\documentclass[twocolumn,times]{aastex63}

\usepackage[caption=false]{subfig}

\def\kms{km~s$^{-1}$}
\def\kmsmpc{km s$^{-1}$ Mpc$^{-1}$}

\def\ergs{erg s$^{-1}$}

\def\msun{\ifmmode M_{\odot} \else $M_{\odot}$\fi}
\def\msunyr{\ifmmode M_{\odot}~{\rm yr}^{-1} \else M$_{\odot}$~yr$^{-1}$\fi}
\def\zsun{\ifmmode Z_{\odot} \else Z$_{\odot}$\fi}
\def\lsun{\ifmmode L_{\odot} \else L$_{\odot}$\fi}

\newcommand{\mstar}{\ifmmode M_\star \else $M_\star $\fi}
\newcommand{\luv}{\ifmmode L_{\rm UV} \else $L_{\rm UV}$\fi}
\newcommand{\lir}{\ifmmode L_{\rm IR} \else $L_{\rm IR}$\fi}
\newcommand{\lbol}{\ifmmode L_{\rm bol} \else $L_{\rm bol}$\fi}

\received{xxx}
\revised{yyy}
\accepted{zzz}
\shorttitle{Ionized vs molecular gas velocity dispersion in galaxies}
\shortauthors{Girard et al.}

\begin{document}

\title{Systematic difference between ionized and molecular gas velocity dispersion in $z\sim1-2$ disks and local analogues}
\correspondingauthor{Marianne Girard}
\email{mgirard@swin.edu.au}

\author[0000-0002-8583-2521]{M. Girard} 
\affiliation{Centre for Astrophysics and Supercomputing, Swinburne University of Technology, P.O. Box 218, Hawthorn, VIC 3122, Australia}
\affiliation{ARC Centre of Excellence for All Sky Astrophysics in 3 Dimensions (ASTRO 3D), Australia}

\author[0000-0003-0645-5260]{D. B. Fisher}
\affiliation{Centre for Astrophysics and Supercomputing, Swinburne University of Technology, P.O. Box 218, Hawthorn, VIC 3122, Australia}
\affiliation{ARC Centre of Excellence for All Sky Astrophysics in 3 Dimensions (ASTRO 3D), Australia}

\author[0000-0002-5480-5686]{A. D. Bolatto}
\affiliation{University of Maryland 1113 PSC Bldg., 415 College Park, MD 20742-0001, USA}

\author[0000-0002-4542-921X]{R. Abraham}
\affiliation{Department of Astronomy \& Astrophysics, University of Toronto, 50 St. George Street, Toronto, ON M5S 3H4, Canada}

\author[0000-0001-9760-7519]{R. Bassett} 
\affiliation{Centre for Astrophysics and Supercomputing, Swinburne University of Technology, P.O. Box 218, Hawthorn, VIC 3122, Australia}
\affiliation{ARC Centre of Excellence for All Sky Astrophysics in 3 Dimensions (ASTRO 3D), Australia}

\author[0000-0002-3254-9044]{K. Glazebrook}
\affiliation{Centre for Astrophysics and Supercomputing, Swinburne University of Technology, P.O. Box 218, Hawthorn, VIC 3122, Australia}
\affiliation{ARC Centre of Excellence for All Sky Astrophysics in 3 Dimensions (ASTRO 3D), Australia}

\author[0000-0002-2775-0595]{R. Herrera-Camus}
\affiliation{Astronomy Department, Universidad de Concepci\'on, Barrio Universitario, Concepci\'on, Chile}

\author[0000-0003-2517-4931]{E. Jim\'enez}
\affiliation{International Centre for Radio Astronomy Research (ICRAR), M468, University of Western Australia, 35 Stirling Hwy., Crawley, WA 6009, Australia}
\affiliation{ARC Centre of Excellence for All Sky Astrophysics in 3 Dimensions (ASTRO 3D), Australia}

\author[0000-0003-4023-8657]{L. Lenki\'c}
\affiliation{University of Maryland 1113 PSC Bldg., 415 College Park, MD 20742-0001, USA}

\author[0000-0002-1527-0762]{D. Obreschkow}
\affiliation{International Centre for Radio Astronomy Research (ICRAR), M468, University of Western Australia, 35 Stirling Hwy., Crawley, WA 6009, Australia}




\begin{abstract}
We compare the molecular and ionized gas velocity dispersion of 9 nearby turbulent disks, analogues to high-redshift galaxies, from the DYNAMO sample using new ALMA and GMOS/Gemini observations. We combine our sample with 12 galaxies at $z\sim $0.5-2.5 from the literature. We find that the resolved velocity dispersion is systematically lower by a factor $2.45\pm0.38$ for the molecular gas compared to the ionized gas, after correcting for thermal broadening. This offset is constant within the galaxy disks and indicates the co-existence of a thin molecular and thick ionized gas disks. This result has a direct impact on the Toomre $Q$ and pressure derived in galaxies. We obtain pressures $\sim0.22$ dex lower on average when using the molecular gas velocity dispersion, $\sigma_{0,mol}$. We find that $\sigma_{0,mol}$ increases with gas fraction and star formation rate. We also obtain an increase with redshift and show that the EAGLE and FIRE simulations overall overestimate $\sigma_{0,mol}$ at high redshift. Our results suggest that efforts to compare the kinematics of gas using ionized gas as a proxy for the total gas may overestimate the velocity dispersion by a significant amount in galaxies at the peak of cosmic star formation. When using the molecular gas as a tracer, our sample is not consistent with predictions from constant efficiency star formation models, even when including transport as a source of turbulence. Feedback models with variable star formation efficiency, $\epsilon_{ff}$, and/or feedback efficiency, $p_*/m_*$, better predict our observations. 
\end{abstract}



\section{Introduction} 
\label{sec:intro}

The majority of the stellar mass in the Universe was built up around $z\sim2$, when the cosmic star formation rate density reached its maximum \citep{Madau2014}. Many surveys have therefore targeted galaxies at this epoch to study their morphology and kinematic properties to gain information on the processes driving galaxy evolution \citep[e.g.,][]{Stott2013,Wisnioski2015, Swinbank2017}.

Important differences between local and high-redshift galaxies have been reported from these surveys. The distant systems present a clumpy morphology \citep[e.g.][]{Elmegreen2005, Genzel2011} and show a higher gas fraction than local galaxies \citep[e.g.][]{Tacconi2013, Dessauges-Zavadsky2015}. Moreover, most of the $z\sim1-2$ galaxies show rotating disks \citep[e.g.][]{ForsterSchreiber2009, Wisnioski2015, Stott2016}, similarly to local disk galaxies, but they also have a much higher velocity dispersion on average with values of $\sigma_0\sim50$ \kms \, \citep[e.g.][]{Johnson2018, Girard2018a, Ubler2019}. This high velocity dispersion has been interpreted as turbulence in the disk of galaxies and could possibly be the result of star formation feedback \citep[e.g.][]{Lehnert2009, Green2010}. Gravitational instabilities in the disk of gas-rich galaxies could also play a major role \citep[e.g.][]{Bournaud2010, Ceverino2010}. \citet{White2017} found a relationship between the kinematic properties and gas fraction of galaxies consistent with predictions from the Toomre disk instability theory ($Q\sim1$). \citet{Fisher2017a} also found that the clump properties in these turbulent disks are in good agreement with this theory. Recent models combining stellar feedback and gravitational instabilities have also successfully predicted the relation between the velocity dispersion and star formation rate at this epoch \citep{Krumholz2018, Ubler2019}.

Most of these studies used strong emission lines, which trace the ionized gas, to obtain the galaxy kinematics. They, therefore, assumed that this component shows similar kinematic properties to the molecular gas, more difficult to detect in distant galaxies. It is now possible to resolve the CO emission and obtain the molecular gas kinematics in some galaxies at $z>1$ with ALMA, often with the help of gravitational lensing \citep[e.g.,][]{Swinbank2011,Barro2017, CalistroRivera2018, Ubler2018, Girard2019, Molina2019}. The molecular and ionized gas kinematics have been compared in only a few objects. \citet{Ubler2018} reported a galaxy at $z\sim1.4$ with similar velocity dispersion for the molecular and ionized gas ($\sigma_0\sim15-30$ \kms). \citet{Girard2019} also found  similar velocity dispersions for one galaxy at $z\sim1$ ($\sigma_0\sim19$ \kms), but they also obtained a discrepancy for a spiral galaxy at $z\sim1$ with a lower molecular velocity dispersion ($\sigma_{0,mol}\sim11$ \kms \, and $\sigma_{0,ion}\sim54$ \kms). More observations are needed to establish the relation between the molecular and ionized gas velocity dispersion at $z\sim1-2$.

In this work, we use DYNAMO \citep{Green2014, Fisher2017a}, a sample of rare nearby galaxies at $0.075<z<0.2$ showing similar physical properties to $z\sim1-2$ main sequence galaxies, to probe the molecular and ionized gas kinematics. By combining our new ALMA observations with our GMOS/Gemini observations, we compare the molecular and ionized gas velocity dispersion of 9 galaxies at a spatial resolution of $\sim1.2$~kpc.

The paper is organized as follows. Section \ref{Sect:sample} describes the sample selection, observations and analysis. In Sect. \ref{Sect:results} and \ref{sec:4}, we present our results and discussion on the relation between the kinematics properties and other physical properties of the galaxies. We finally present a summary in Sect. \ref{Sect:conclusion}. 

In this paper, we adopt a $\Lambda$-CDM cosmology with $H_0=67$ \kmsmpc , $\Omega_M=0.31$, and $\Omega_\Lambda=0.69$. We adopt a \citet{Kroupa2001} initial mass function (IMF) and convert to this IMF when appropriate.

\begin{center}
\begin{deluxetable*}{  m{2.6cm}  m{1.1cm} m{1cm} m{1.4cm} m{1.7cm} m{1cm} m{1cm} m{1.2cm} m{1.5cm}  m{2.9cm} } 
\caption{Galaxy properties\label{tab:properties}}
\tablewidth{0pt}
\tablehead{
Galaxy & \colhead{z} & \colhead{\mstar} & \colhead{SFR} & \colhead{$f_{gas} \, ^a$} & \colhead{$\upsilon_{rot,ion} \, ^{b,c}$} & \colhead{$\upsilon_{rot,mol} \,^{c}$} & \colhead{$\sigma_{0,ion} \, ^{b}$}& \colhead{$\sigma_{0,mol}$} & \colhead{Reference}\\
\colhead{} & \colhead{} & \colhead{[$10^{10}$\msun]} & \colhead{[\msunyr]} & \colhead{} & \colhead{[\kms]} & \colhead{[\kms]} & \colhead{[\kms]}& \colhead{[\kms]} 
}
\startdata
C13-1 & 0.07876 & 3.58 & $5.1 \pm 0.5$ & $0.06 \pm 0.02$ & 223 & 234 & 26$\pm4$ & 8$\pm2$ & This work \\
C22-2 & 0.07116 & 2.19 & $3.2 \pm 0.2$  &  $0.18\pm 0.03$ & 191 & 208 & 35.5$\pm5$ & 9.5$\pm3$ & This work \\
D13-5 & 0.07535 & 5.38 & $17.5 \pm 0.5$  & $0.36\pm0.02$ & 175 & 190 & 39.5$\pm5$ & 12$\pm3$ & This work \\
D15-3 & 0.06712 & 5.42 & $ 8.3 \pm 0.4$ &  $0.17\pm0.04$ & 244 & 230 & 25$\pm5$ & 8$\pm2$ & This work \\
G04-1 & 0.12981 & 6.47 &  $ 21.3 \pm 1$ & $0.33\pm0.04$ & 274 & 248 &  30$\pm5$ & 13$\pm3$ & This work \\
G08-5 & 0.13217 & 1.73 &   $ 10.0 \pm 1$ & $0.30\pm0.05$ & 214 & 227 & 36$\pm7$ & 15$\pm4$ & This work \\
G14-1 & 0.13233 & 2.23 &  $ 6.9 \pm 0.5$ & $0.77\pm0.08$ & 149 & 160 & 71$\pm9$ & 27$\pm5$ & This work \\
G20-2 & 0.14113 & 2.16 &  $18.2 \pm 0.4$ & $0.21\pm0.05$ & 151 & 160 & 36$\pm5$ & 9$\pm4$ & This work \\
SDSS 013527-1039 & 0.127 & 7.08 &  $ 25.3 \pm 2.5$ & $0.33\pm0.04$ & 172 & 206 & 49$\pm7$ & 12.5$\pm4$ & This work \\
\hline
SMM J2315-0102  & 2.32 & 2.01 &  $268$ & $0.88\pm 0.10$ & - & 320 & - & $40\pm10$ & \citet{Swinbank2011} \\
HLS115   &  1.58 & 2.63 &  $120\pm6 $ & $0.75\pm0.15$ & 218 & - & $80\pm10$ & - & \citet{Girard2018b}\\
Cosmic Snake  & 1.036 & 2.51 &  $  19\pm6 $ & $0.29\pm0.05$ & 225 & - & 18.5$\pm7$ & 19.5$\pm6$ &\citet{Patricio2018}, \citet{Girard2019} \\
A521 & 1.044 & 3.41 &  $ 9.4\pm3.0 $ & - & 130 & - & $54\pm11$  & $11\pm8$ &\citet{Patricio2018}, \citet{Girard2019} \\
SHiZELS-19 & 1.486 & 2.12 &  $13.8\pm$2.0 & $0.82\pm0.16$ & 106 & 121 & $107\pm13$  & $91\pm6$ & \citet{Molina2019}\\
G3\_10098 & 0.66 & 13.4 &  $ 53 $ & $0.19$ & - & - & - & $39.8\pm15.6$ & PHIBSS$^{d}$\\
G4\_21351 & 0.73 & 8.45 &  $ 85 $ & $0.32$ & -& - & - & $19.5\pm5.8$ & PHIBSS$^{d}$\\
EGS\_13035123 & 1.12 & 16.8  &  $ 85 $ & $0.32$ & - & 193 & - & $20.4\pm5.2$ & PHIBSS$^{d}$\\
EGS\_12007881 & 1.17 & 5.33 &  $106 $ & $0.54$ & - & 220 & - & $22.9\pm5.8$  & PHIBSS$^{d}$\\
EGS\_13003805 & 1.23 & 16.9 &  $212 $ & $0.48$ &-  & 384 & - & $30.2\pm8.3$ & PHIBSS$^{d}$\\
EGS4\_24985 & 1.4 & 8.45 &  $106 $ & $0.37$ & - & - & - & $19.0\pm7.0$ & PHIBSS$^{d}$\\
EGS\_13011166 & 1.53 & 13.4 &  $ 423 $ & $0.65$ & - & 342 & - & $33.1\pm7.6$ & PHIBSS$^{d}$\\
\enddata
\tablecomments{\\
$^a$ \, Values from \citet{Fisher2017a}, \citet{White2017} and \citet{Fisher2019} for the DYNAMO galaxies. \\
$^b$ \, We do not have GMOS/Gemini data for C13-1 and D15-3. The ionized gas kinematic properties for these targets were taken from \citet{Bekiaris2016} which used a very similar kinematic modeling method. Their data is however at lower resolution. \\
$^c$ \, Uncertainty on the rotation velocity of the ionized gas, $\upsilon_{rot,ion}$, and molecular gas, $\upsilon_{rot,mol}$, of the DYNAMO galaxies is $\sim$15 \kms. \\
$^d$ \, The properties of the PHIBSS galaxies have been taken from \citet{Genzel2010},\citet{Tacconi2010,Tacconi2013}, \citet{Freundlich2013}, \citet{Genzel2015}, \citet{Tacconi2018}, \citet{Ubler2018,Ubler2019}, and \citet{Freundlich2019}.
}
\end{deluxetable*}  
\end{center}

\section{Sample selection, observations and analysis}
\label{Sect:sample}

\subsection{Sample}

The original sample of 67 DYNAMO galaxies in \citet{Green2010,Green2014} includes a heterogenous mix of targets. This sample has been down-selected to a sample of $\sim$10 targets in which a wide range of properties are shown to match with $z\approx1.5$ main-sequence galaxies. The subset of these ``$z=1.5$ analogues" galaxies is presented in this work and \citet{Fisher2017a, Fisher2019}.
Our sample includes galaxies at redshift $0.075<z<0.2$ with a stellar mass of $1\times10^{10}<\mstar[\msun]<9\times10^{10}$ and star formation rate of $1<$SFR [\msunyr]$<60$. These galaxies have therefore a similar specific star formation rate to main-sequence galaxies at $z\sim1-2$. They also have a high gas fraction $f_{gas}\sim0.1-0.8$ \citep{Fisher2014, White2017,Fisher2017a} and a clumpy morphology \citep{Fisher2017b}, typical of galaxies at $z\sim1-2$ \citep[e.g.][]{Tacconi2013, Dessauges-Zavadsky2015, Tacconi2018}. The kinematic properties of the DYNAMO galaxies also present similarities with high-redshift galaxies. DYNAMO galaxies show a rotating disk, but are also turbulent, with a high ionized gas velocity dispersion $\sigma_{0,ion}\sim 20-100$ \kms \, \citep{Green2014, Bassett2014, Bekiaris2016, Oliva-Altamirano2018}. They also have a low angular momentum \citep{Obreschkow2015}, as found for galaxies with similar mass at $z\sim1-2$ \citep{Swinbank2017}.

To study the molecular gas and ionized gas of the galaxies presented in this paper, we use ALMA observations for all 9 galaxies and GMOS/Gemini observations for 7 galaxies  \citep{Bassett2014} as well as SPIRAL/AAT observations for the two other \citep{Green2014}.

\subsection{Observations}

ALMA observations were carried out in 2018 (project 2017.1.00239.S - PI: D. B. Fisher). The total on-source time varies between 5 min and 22 min for the CO(3-2) line at a frequency of 290-310 GHz and between 20 min and 107 min for the CO(4-3) line at a frequency of 390-420 GHz. The data reduction was performed using the standard pipeline from the ALMA facility. The final beam sizes are 0.8''$\times$0.9'' and 0.6''$\times$0.75'' (or 1.8$\times$2.0 kpc and 1.3$\times$1.7 kpc), and the spectral resolutions are $\sim15$ \kms \, and $\sim11$ \kms \, for the CO(3-2) and CO(4-3) lines, respectively.

We also use in this work deep CO(3-2) data of the galaxies G04-1, G08-5, and G14-1 obtained in 2019 (project \#2019.1.00447.S - PI: R. Herrera-Camus). The total on-source time was between 83 and 94 min. Similarly to our previous data, we use the standard pipeline from the ALMA facility for the data reduction. The final beam size is about $0.37"\times0.31"$ with a spectral resolution of $\sim 8$ \kms. 

Observations from GMOS/Gemini were carried out in 2011-2012. The average seeing for these observations is $\sim0.5''$ and the spectral resolution is $\sigma=24$ \kms. The physical distance resolution is $\sim1.1$ kpc. For more details on the observations and data reduction see \citet{Bassett2014}.

\begin{figure*}[ht]
\includegraphics[width=\linewidth]{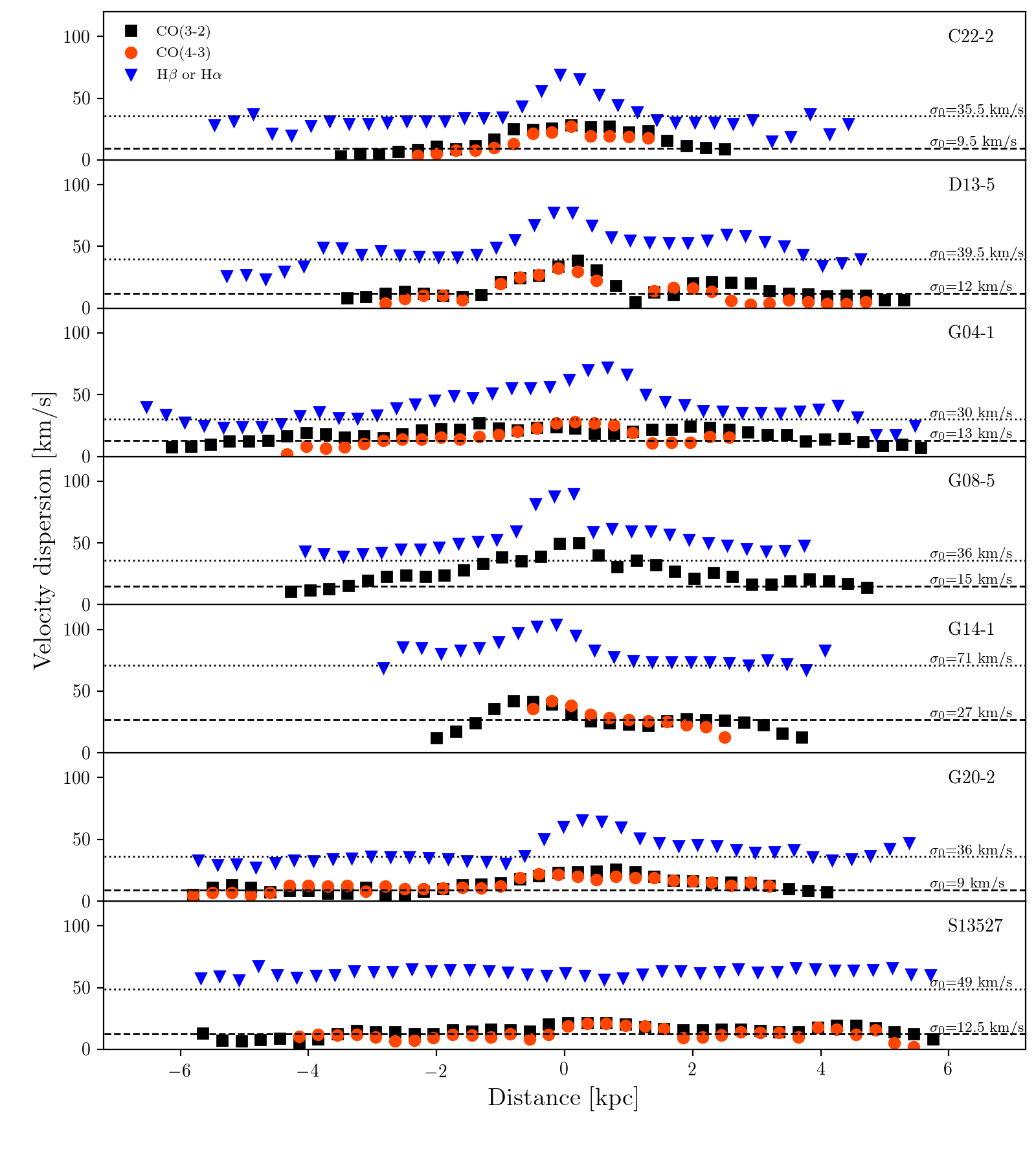}
\caption{Velocity dispersion corrected for beam-smearing as a function of the distance from the kinematic center averaged in bins of 300 pc. The H$\alpha$ or H$\beta$ velocity dispersion profiles obtained from GMOS/Gemini observations are shown as blue triangles. The CO(4-3) and CO(3-2) velocity dispersions obtained from ALMA observations are shown as dark orange circles and black squares, respectively. The dashed and dotted horizontal lines indicate the intrinsic velocity dispersion obtained with GalPak$\rm^{3D}$ for the molecular and ionized gas, respectively. Uncertainty on the ionized and molecular gas velocity dispersion are 5-10 \kms \, and 3-5 \kms, respectively, on average in each bin.
\label{fig:disp_profile}}
\end{figure*}

\subsection{Measurements}

\subsubsection{Gas mass}

The total gas mass and gas fraction of each galaxy were derived in \citet{Fisher2014}, \citet{White2017}, and \citet{Fisher2017a, Fisher2019} using CO(1-0) observations from Plateau de Bure.
The molecular gas mass was obtained using $M_{mol}=\alpha_{CO} L_{CO(1-0)}$, where $L_{CO(1-0)}$ is the CO(1-0) luminosity and $\alpha_{CO}$ is the  CO-to-H$_2$ conversion factor of 4.36, which includes a factor 1.36 for helium. We define the gas fraction as $f_{gas}= M_{mol}/(M_{mol}+\mstar)$ in this work. 

\subsubsection{Star formation rate}

The star formation rates have also been derived in previous DYNAMO papers using the H$\alpha$ flux corrected for attenuation \citep{Green2017,Fisher2019}, where SFR [\msunyr] $=5.53\times10^{-42} L_{H\alpha}$ [\ergs] \citep{Hao2011}. Values are shown in Table \ref{tab:properties} and assume a Kroupa IMF \citep{Kroupa2001}.

\subsubsection{Kinematics}

We obtain the moment 1 and 2 maps from CO(3-2) and CO(4-3) using the \textit{immoments} routine in CASA with a 3$\sigma$ threshold parameter above the noise level. To obtain the kinematic maps from the ionized gas, we use the same method as described in \citet{Bassett2014} and perform Gaussian fits on H$\alpha$ or H$\beta$ in individual spaxel to obtain the flux, velocity and velocity dispersion maps. The instrumental broadening, $\sigma_{instr}=24$ \kms, was subtracted in quadrature to every spaxel of the velocity dispersion maps. The kinematic maps are shown in Figure \ref{map_galaxies}.

We model the galaxy kinematics using GalPaK$\rm^{3D \,} $\footnote{\url{http://www.ascl.net/1501.014}} \citep{Bouche2015}, a 3D galaxy disk modeling code that fits directly a model on the datacubes using a Markov-Chain Monte Carlo method. The model is convolved with the point-spread function (PSF) and line-spread function (LSF). In this way, the beam-smearing effect is directly taken into account during the fitting. We adopt an arctan function to model the velocity \citep{Courteau1997} and an exponential profile for the radial flux profile. The disc inclinations are fixed using the SDSS r-band photometry, similarly to \citet{Bekiaris2016}. This model assumes a constant velocity dispersion over the galaxy disk. The galaxy center, disk half-light radius, the flux, the position angle, the turnover radius, the maximum rotation velocity, and the intrinsic velocity dispersion are then free to vary. We use both CO(3-2) and CO(4-3) datacubes to model the molecular gas kinematics, except for C13-1 and G08-5 for which only CO(3-2) was available and D15-3 for which only CO(4-3) was available. We obtain similar velocity dispersion values for both CO transitions ($\pm 2$ \kms). For the ionized gas, we fit the H$\beta$ line, except for SDSS-13527 for which we use H$\alpha$. Table \ref{tab:properties} shows the intrinsic velocity dispersion, $\sigma_0$, and the inclination-corrected rotation velocity, $\upsilon_{rot}$, obtained from GalPaK$\rm^{3D}$. No GMOS/Gemini data were available for C13-1 and D15-3. For these two galaxies, we take the values from \citet{Bekiaris2016} that use a very similar modeling method with observations from AAT \citep{Green2014}.

To obtain beam-smearing corrected velocity dispersion profiles of each galaxy, we subtract in quadrature the unresolved velocity shear obtained from our models with GalPak$\rm^{3D}$ to the observed velocity dispersion map, already corrected for instrumental broadening \citep{Epinat2012,Bassett2014,Green2014,Levy2018}. This correction affects mostly the central part of the profile, where the velocity gradient is at its steepest. We note that adopting a different beam-smearing correction could  make the central part of the dispersion profile flatter or steeper, but this would not affect significantly the values in the disk. Fig. \ref{fig:disp_profile} presents the velocity dispersion profiles of the 7 galaxies for which both the ionized and molecular gas maps were available. 

The intrinsic velocity dispersions obtained from GalPaK$\rm^{3D}$, also shown in Fig. \ref{fig:disp_profile}, are in good agreement with measurements taken on the major axis in the outer region on the velocity dispersion maps, where the beam-smearing is known to be negligible \citep{Burkert2016, Johnson2018}. The values are consistent at $\pm3$ \kms \, and $\pm5$ \kms \, on average for the molecular and ionized gas, respectively. Our ionized gas velocity dispersions are also in good agreement ($\pm~5$ \kms) with previous results obtained for the ionized gas by \citet{Bassett2014} and \citet{Bekiaris2016} using a very similar technique to model the kinematics.


\begin{figure*}[ht]
\includegraphics[width=\linewidth]{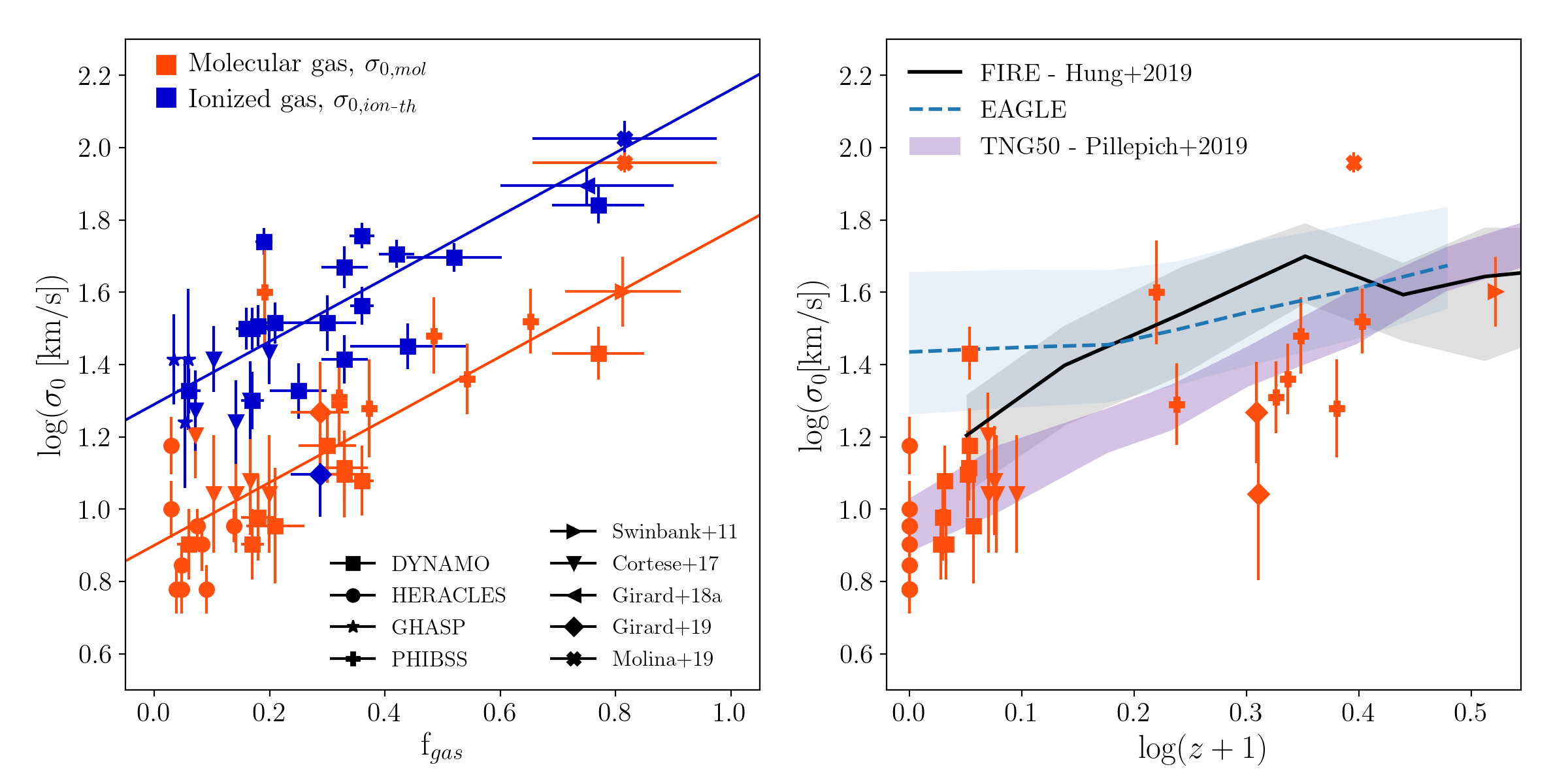}
\caption{Intrinsic velocity dispersion, $\sigma_0$,  as a function of gas fraction, $f_{gas}$ and redshift, $z$. The ionized gas velocity dispersion is corrected for thermal broadening using $\sigma_{0,ion\text{-}th} = (\sigma_{0,ion}^2-\sigma_{th}^2 )^{1/2}$, where $\sigma_{th}=15$ \kms. The squares represent the DYNAMO galaxies from this work and from \citet{Fisher2019}. We also present galaxies at $z>1$ from \citet{Swinbank2011}, \citet{Girard2018b}, \citet{Molina2019}, and \citet{Girard2019} and the PHIBSS survey \citep{Genzel2010,Tacconi2010,Tacconi2013, Freundlich2013, Genzel2015, Tacconi2018, Ubler2018, Ubler2019,  Freundlich2019}. We show local galaxies from the THINGS-HERACLES survey \citep{Leroy2008, Leroy2009, Mogotsi2016}, GHASP \citep{Dunne2000, Leroy2005, Epinat2010, TorresFlores2011} and nearby galaxies at $z\sim0.2$ from \citet{Cortese2017}. The molecular gas is represented in dark orange and the ionized gas in blue. The gas fractions of all galaxies are computed assuming a conversion factor $\alpha_{co}=4.36$. The blue and dark orange lines in the left panel represents a fit to the data of log($\sigma_0$)=0.87$\times f_{gas}+b$ with $b=1.29$ and $b=0.90$ for the ionized and molecular gas, respectively. The black and light blue dashed lines in the right panel represent results from the FIRE \citep{Hung2019} and EAGLE simulations. The grey and light blue area encloses 68\% of the data of the simulated galaxies. The purple area indicates the results from TNG50 \citep{Pillepich2019}. The upper and lower limits are the values for 10.5$<$log(\mstar/\msun)$<$11.0 and 10.0$<$log(\mstar/\msun)$<$10.5 without a thermal contribution. 
\label{fig:disp_fgas}}
\end{figure*}

\section{Results and discussion}
\label{Sect:results}

In this section, we compare the molecular and ionized gas velocity dispersion. We explore the correlations of the velocity dispersion with gas fraction and redshift.

\subsection{Difference between $\sigma_{0,ion}$ and $\sigma_{0,mol}$ and correlation with gas fraction}

Table \ref{tab:properties} shows the main kinematic properties obtained for the ionized and molecular gas of the DYNAMO galaxies. We find that the rotation velocity of the molecular gas is systematically higher than that of ionized gas with an average difference of 10-15 \kms. We do note that the uncertainty in our measurement of the rotation velocity is $\sim15$ \kms. Nonetheless this offset is similar to what is found in local spiral galaxies from the EDGE-CALIFA survey \citep{Levy2018}.

Fig. \ref{fig:disp_profile} presents the velocity dispersion profiles and the intrinsic velocity dispersion values obtained from our kinematic models. We find that the dispersion profiles for both the molecular and ionized gas are relatively flat in the galaxy disks. We also see an increase in the center of the galaxies, potentially due to the presence of a small bulge \citep{Bassett2014}.

We find that at all radii, in all galaxies the molecular gas velocity dispersion is systematically lower than that of ionized gas, with $\sigma_{0,mol}\sim 8-27$ \kms \, and $\sigma_{0,ion}\sim 25-71$~\kms \, for the molecular and ionized gas, respectively. The ionized gas velocity dispersion is expected to be higher due to thermal broadening and the expansion of HII regions. These two effects together can account for $\sigma_{th}\sim15$ \kms \, \citep{Shields1990, Krumholz2016, Krumholz2018}. After correcting in quadrature for this component, we find that the ionized gas dispersion is still higher by a factor 2-3 compared to the molecular gas dispersion and that the difference varies between 12-42 \kms \, in the sample. We also find that the difference does not change within a galaxy for radii outside the central region, meaning that there is no correlation between the difference in velocity dispersion and $\Sigma_{SFR}$, $\Sigma_{mol}$ and $\Sigma_{*}$.

The gas disk thickness can be estimated using $\sigma_0/\upsilon_{rot} \approx H/R$, where $H$ is the height and $R$ is the radius of the disk \citep{Genzel2011, White2017}. In DYNAMO galaxies, we find that $R_{1/2, H_\alpha}= (1.1\pm0.2) \, R_{1/2, CO}$. This suggests that the molecular gas disk of our DYNAMO sample is thinner than the ionized gas disk since the radius of the molecular and ionized gas disks are similar in these galaxies. \citet{Girard2019} also found the co-existence of a thin molecular gas disk and thick ionized gas disk in lensed galaxies at $z\sim1$.

Fig. \ref{fig:disp_fgas} (left panel) shows that the ratio between the ionised and molecular velocity dispersion is nearly constant across a large range in gas fractions.  The ionized velocity dispersions presented in this plot have been corrected in quadrature for the thermal broadening and HII regions expansion ($\sigma_{th}\sim15$ \kms). We also plot DYNAMO galaxies from \citet{Fisher2019}, samples of local galaxies from THINGS-HERACLES \citep{Leroy2008,Leroy2009, Mogotsi2016}, GHASP \citep{Dunne2000, Leroy2005, Epinat2010, TorresFlores2011}, nearby galaxies from \citet{Cortese2017}, and high-redshift galaxies from lensed studies \citep{Swinbank2011, Girard2018b, Girard2019}, from \citet{Molina2019}, and from PHIBSS \citep{Genzel2010,Tacconi2010,Tacconi2013, Freundlich2013, Genzel2015, Tacconi2018, Ubler2018, Ubler2019,  Freundlich2019}. 

The gas fraction for all the galaxies was computed using the Galactic CO-H$_2$ conversion factor $\alpha_{co}=4.36$, except for the PHIBSS sample where they also corrected $\alpha_{CO}$ for metallicity by adopting the geometric mean of the prescriptions presented in \citet{Bolatto2013} and \citet{Genzel2012}. However, the metallicity in massive galaxies at $z\sim1-2$ is not low enough (estimated to $>0.8$ times the solar value for these galaxies) to have a significant effect on $\alpha_{CO}$ \citep{Tacconi2013,Tacconi2020}. On the other hand, the galaxies with a high baryonic surface density could possibly have a lower conversion factor \citep{Narayanan2012, Bolatto2013}, but the relation we present here shows a relatively constant offset between the velocity dispersion, which would be unaffected by the conversion factor. The choice of the conversion factor $\alpha_{CO}$ here is therefore not critical.

We obtain a strong correlation between the molecular velocity dispersion and gas fraction with a Spearman rank correlation coefficient of 0.76. The increase of the velocity dispersion with gas fraction is observed for both the ionized and molecular gas and agrees with the picture of more turbulent, gas-rich galaxies at high-redshift \citep[e.g.,][]{Tacconi2013, Wisnioski2015, Johnson2018}. When fitting the whole sample, we obtain a fit of log($\sigma_0$/\kms) = $a \times f_{gas} + b$ with $a=0.98\pm0.17$ and $a=0.76\pm0.11$ for the molecular and ionised gas, respectively, and $a=0.87\pm0.13$ when combining both molecular and ionized gas values. We fit the same relation with a fixed slope of $a=0.87$ on both gas component separately and obtain 
\begin{equation}
\label{eq:Q}
\mathrm{log} \left( \frac{\sigma_{0,mol} }{\mathrm{km \, s^{-1}} } \right) = 0.87 \times f_{gas} + (0.90\pm0.03)
\end{equation}
for the molecular gas, and
\begin{equation}
\label{eq:Q}
\mathrm{log} \left( \frac{\sigma_{0,ion\text{-}th}}{\mathrm{km \, s^{-1}}} \right) = 0.87 \times f_{gas} + (1.29\pm0.03)
\end{equation}
for the ionized gas (corrected for the thermal broadening). We find a zero offset between the molecular and ionized gas velocity dispersion of $0.39\pm0.06$ dex, which is equivalent to a factor $2.45\pm 0.38$. 

We also find a mean value of $\sigma_{0,ion\text{-}th}/\sigma_{0,mol} =2.44\pm1.01$ when combining all the galaxies of this sample for which both ionised and molecular gas velocity dispersion are available. This includes galaxies with $0.06<f_{gas}<0.82$ and $0.8<$SFR [\msunyr]$<25$ from DYNAMO, \citet{Cortese2017}, \citet{Girard2019} and \citet{Molina2019}. We find that $\sigma_{0,ion\text{-}th}/\sigma_{0,mol}$ does not correlate with gas fraction (Spearman coefficient of 0.10) and star formation rate (Spearman coefficient of 0.22).

\begin{figure*}[ht]
\centering
\includegraphics[width=\linewidth]{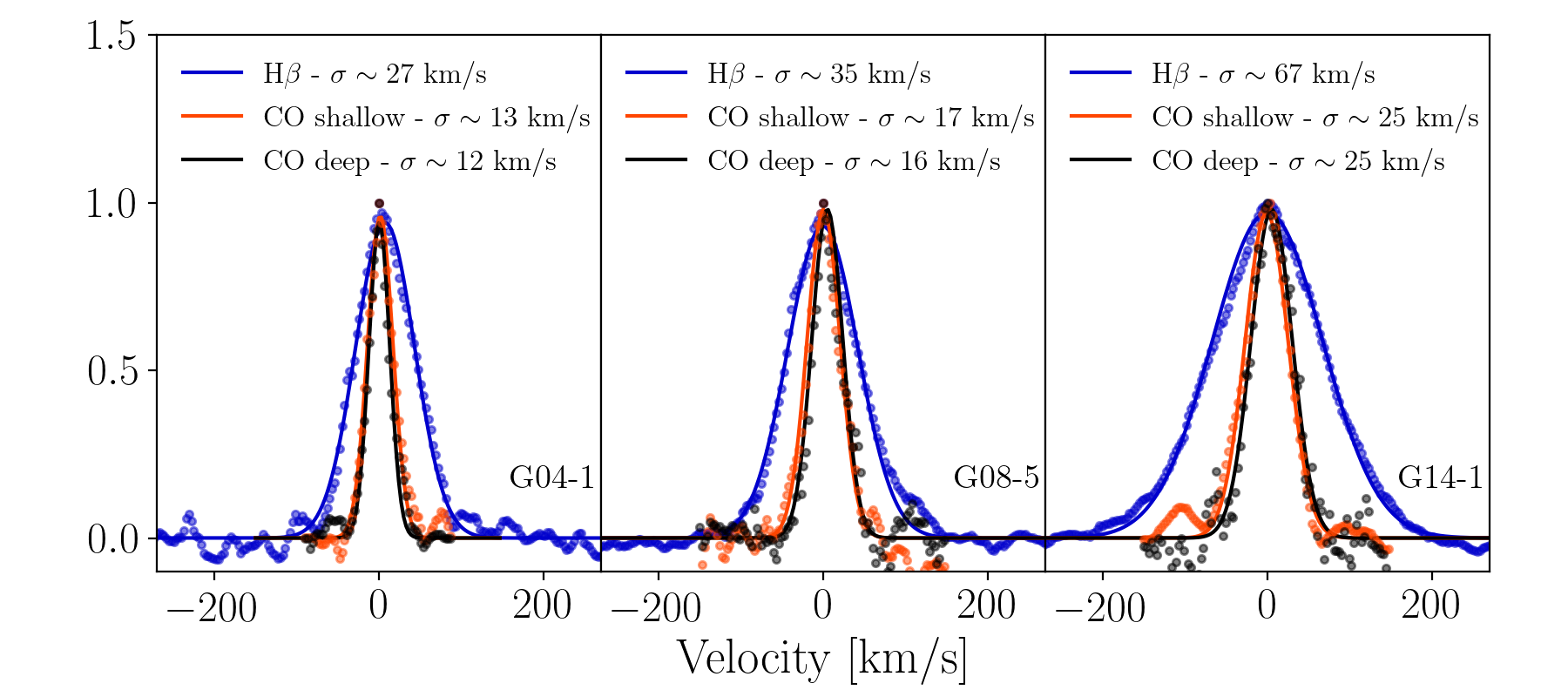}
\caption{Stacked spectra of the H$\beta$ (in blue) and CO(3-2) lines in the outer regions of G04-1, G08-5 and G14-1 for the deep (in black) and shallow data (in dark orange). The solid lines represent a Gaussian fit to the spectra from which we determined the velocity dispersion. The velocity dispersion indicated is corrected for instrumental broadening.}
\label{fig:stack}
\end{figure*}

\citet{Caldu-Primo2013} suggested that deep data are necessary to trace a broader component of the molecular gas velocity dispersion. In Fig.~\ref{fig:stack} we carry out a similar experiment on DYNAMO galaxies, but find this does not account for the difference between ionized and molecular gas dispersions. 
We use deep ALMA CO(3-2) data of the galaxies G04-1, G08-5 and G14-1 obtained in 2019 (PI: R. Herrera-Camus) to compare with the shallower data. To trace a potential broader component, we average spectra from the outer regions of the galaxies for our deep and shallow data \citep[similar method to][]{Ianjamasimanana2012}. To take account of the different velocities in each spaxel, we use the first moment to shift each spectrum to the same velocity. After stacking, we perform a Gaussian fit to the two final spectra to determine the velocity dispersion. Using the Bayesian information criterion (BIC), we also find that a single Gaussian fits is more adequate than a two Gaussian component for these spectra (difference of BIC higher than 10). 
The deep CO stacked data does not show a broader component compared to the shallower stacked CO data,  which is different than what is found in local spirals. This suggests that our sample of turbulent, gas-rich galaxies do not have an additional thicker molecular gas disk. We note however that the galaxy sample in \citet{Caldu-Primo2013} is different. They targeted local spirals from THINGS, which have a primary velocity dispersion of 5-10 \kms \, and a broad secondary component of $\sim$10-20 \kms. The broad component of the THINGS sample is closer to the main component of what we measure in DYNAMO. They also observe the CO(2-1) or CO(1-0) transition.
We also show the results for the H$\beta$ line using the same method for comparison in Fig. \ref{fig:stack}. 

Another possible explanation could be that multiple components are present in the ionized gas. If we interpret the molecular gas as being representative of a narrow component to the gas emission this implies that there is a secondary component heating the ionized gas. There is evidence for this in the literature. \citet{Ho2014} carried out multi Gaussian fits to spiral galaxies with R$\sim$5000 spectra. They found that second components with $\sigma\sim40$ \kms \, are common. Based on line ratios they argued that the ionized gas is heated by both shocks and photoionization. We cannot say from our data if the wider H$\beta$ profiles have similar lines ratios in DYNAMO galaxies, though we note the dispersion is similar. More work at high spectral resolution is needed to determine the nature of the wide emission lines in DYNAMO galaxies.

\citet{Hollenbach1980} also argued that the presence of fast shocks with $\upsilon_{shock}>25$ \kms \, could dissociate the H$_2$ molecules (or $\upsilon_{shock}>50$ \kms \, in low density environment). This could leave a highly turbulent gas component that is not traced by the molecular gas. \citet{Wong2002} suggested that the formation of the molecular gas could be limited to the midplane only due to low pressure or low dust abundance outside this region. Clearly, more studies on the molecular gas kinematics in the limit of high gas surface densities, similar to DYNAMO and high-redshift galaxies, are needed to establish the nature of this important difference.

\subsection{Redshift evolution of $\sigma_{0,mol}$}

Fig. \ref{fig:disp_fgas} (right panel) shows the molecular gas velocity dispersion as a function of redshift for the same sample presented in the left panel. We find that the molecular gas velocity dispersion is higher in high-redshift galaxies, similarly to \citet{Ubler2019}. 

We compare our observations to the results from three different simulations: FIRE \citep{Hung2019}, TNG50 \citep{Pillepich2019}, and EAGLE (Jimenez et al. in prep). 
The grey area shows the results from the FIRE simulation, presented in Table 1 of \citet{Hung2019}, and encloses 68\% of their data. They studied only star-forming galaxies and defined the velocity dispersion as the minimum value across $10^4$ viewing angles (equivalent to a face-on viewing angle for disk galaxies) of the SFR-weighted standard deviation of the velocity distribution of the gas particles. 
The blue area encloses 68\% of the data of the EAGLE simulation. EAGLE includes both star-forming and quenched galaxies with log(\mstar/\msun)$>9$.
In this simulation, the velocity dispersion is defined as the mass-weighted vertical velocity dispersion of the star-forming gas, without any thermal contribution. We use measurements without a thermal component since we compare to cold molecular gas observations. 
Finally, we show in purple the velocity dispersion from TNG50 \citep[Fig. 12  of ][]{Pillepich2019} without a thermal term for a stellar mass between 10.0$<$log(\mstar/\msun)$<$11.0. In their work, they include only star-forming galaxies and defined the velocity dispersion as the unweighted mean pixel-based (0.5 kpc) velocity dispersion of the star-forming gas. We note that the star-forming gas traces both molecular and ionized gas without distinguishing between them, but we do not include here the thermal contribution to compare with the cold gas component. They measured the dispersion on face-on galaxies and average only pixels in the outer region of the galaxies that are on the kinematic major axis, where the rotation curve is flat.

We find that the results from the simulations are overall higher than our molecular gas observations. The values from TNG50 are however lower than FIRE and EAGLE, and more similar to our observations. This could be due to the different methods used to determine the velocity dispersion in the  simulations. FIRE and EAGLE include the velocity dispersion from the center of the galaxies and weight their values using the mass or SFR, while TNG50 only average the dispersion of the pixels in the outer region. From our DYNAMO galaxies, we find that the velocity dispersion is on average 5-10 \kms \, higher when we include the center of galaxies and weight with SFR. This could also be due to different physical processes included in the simulations and the way they are implemented. In TNG50, unlike FIRE, the stellar feedback is hydrodynamically decoupled from the interstellar medium \citep{Pillepich2019, Ubler2020}. The AGN feedback in TNG50 is however directly coupled to the gas. FIRE does not have any AGN feedback. More details about what drives the differences in velocity dispersion in the simulations will be provided in Jimenez et al. (in prep).

\begin{figure*}[ht]
\centering
\subfloat{\includegraphics[width=0.5\textwidth]{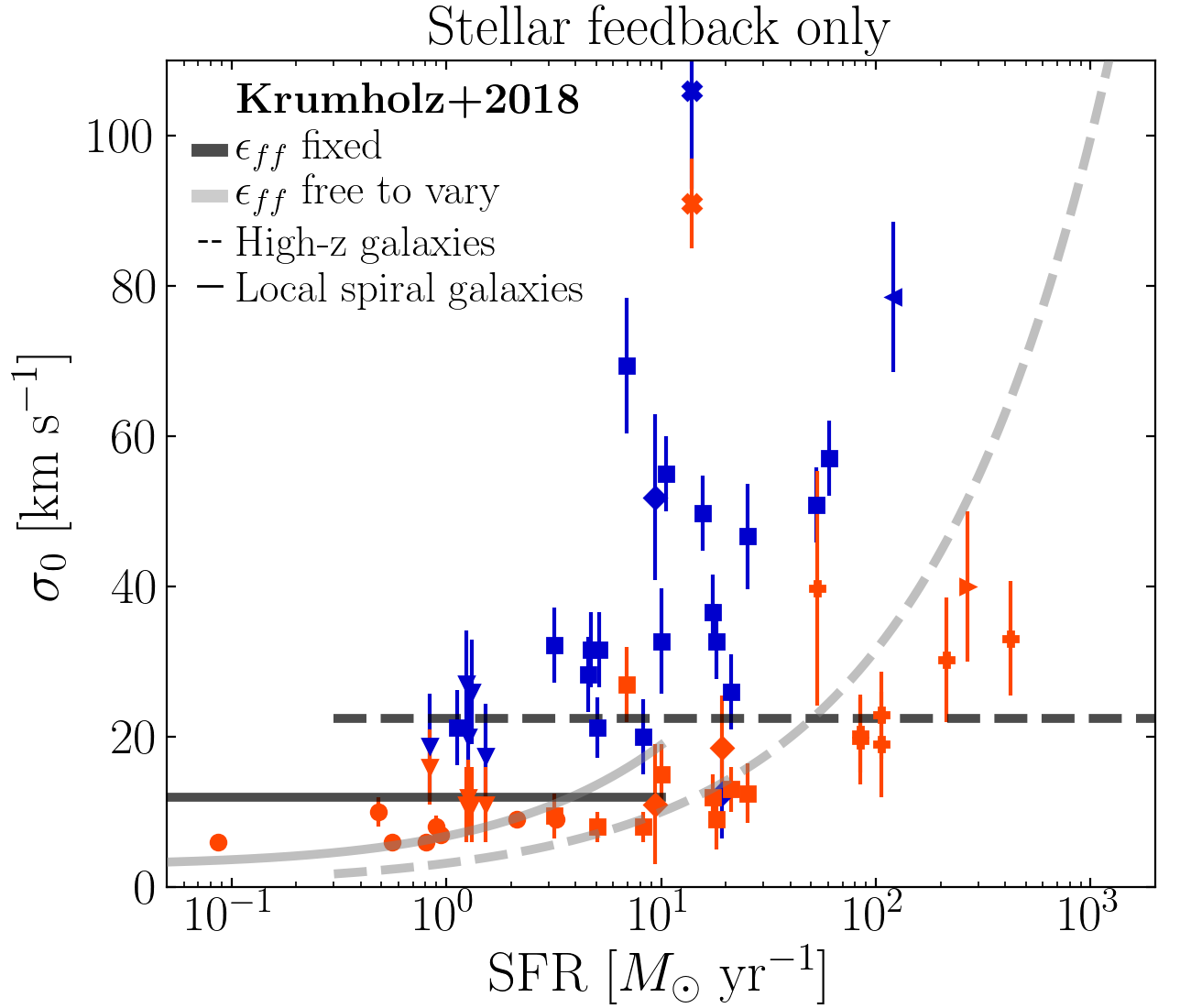}} 
\subfloat{\includegraphics[width=0.5\textwidth]{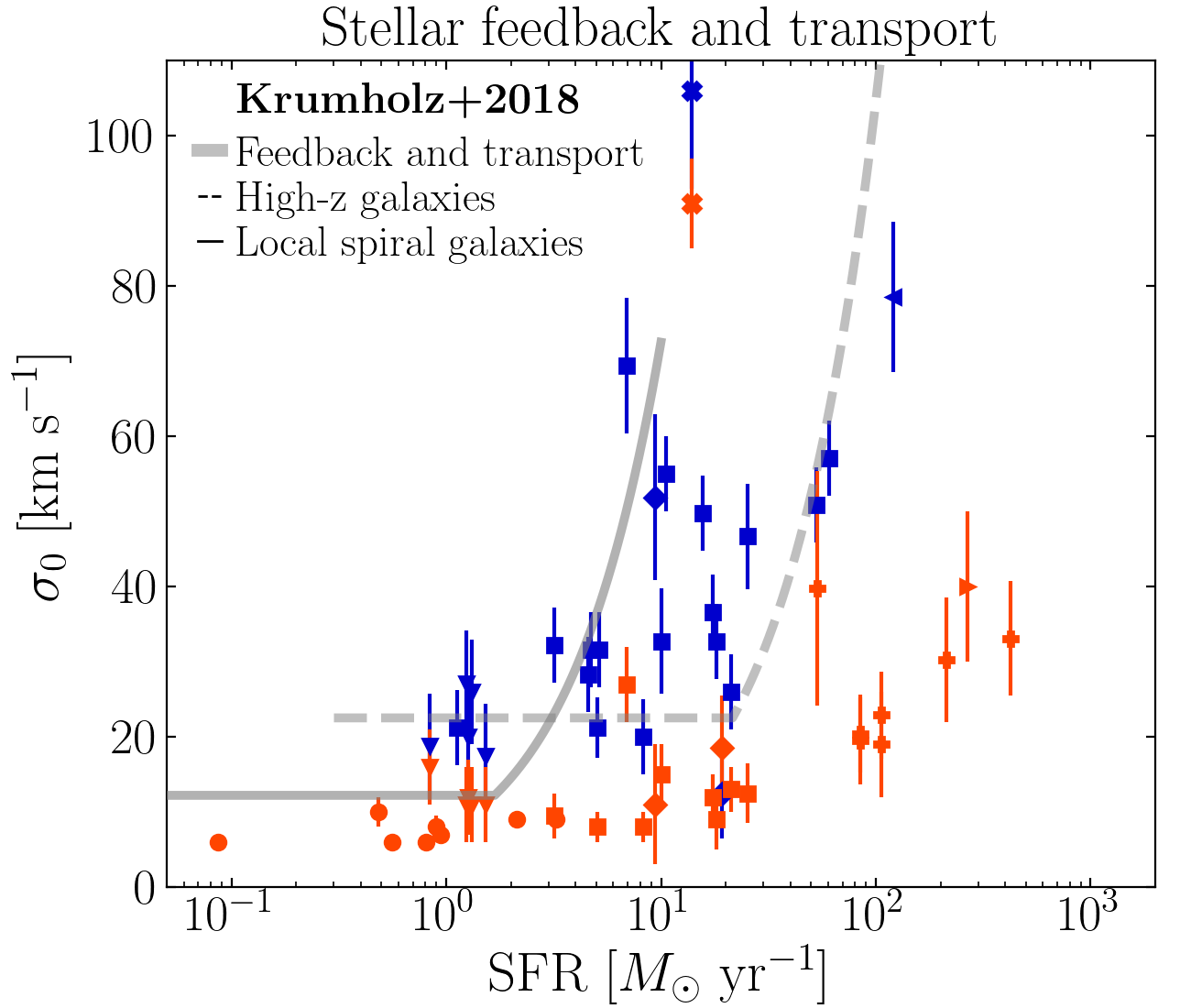}}
\caption{Intrinsic velocity dispersion of the molecular gas, $\sigma_{0,mol}$ (in dark orange), and ionized gas corrected for thermal broadening, $\sigma_{0,ion\text{-}th}$ (in blue), as a function of star formation rate. The same data as in Fig. \ref{fig:disp_fgas} are presented. The left and right panels present the feedback model with  $\epsilon_{ff}$ fixed and free to vary and the model with feedback and gravitational instabilities (transport) from \citet{Krumholz2018}, respectively, using the same fiducial values.} 
\label{fig:disp_sfr}
\end{figure*}

\subsection{Impact on Toomre Q}
\label{sect:toomreQ}

The Toome $Q$ stability parameter \citep{Toomre1964} for the gas component can be expressed as $Q_{gas} = \sigma_0 \kappa/ \pi G \Sigma_{g}$, where $\kappa$ is the epicyclic frequency and $\Sigma_{g}$ is the gas surface density. We derive $Q$ in three different ways: (1) with $\sigma_{0,ion}$ and assuming a fixed $\alpha_{CO}=4.36$; (2) with $\sigma_{0,mol}$ and a fixed $\alpha_{CO}=4.36$; (3) with $\sigma_{0,mol}$ and a variable $\alpha_{CO}(\Sigma_{CO})$. To determine $\alpha_{CO}(\Sigma_{CO})$, we use eq. 31 in \citet[][]{Bolatto2013} assuming a solar metallicity. We use $\alpha_{CO}=4.36$ to initially estimate $\Sigma_{CO}$, and then recalculate $\Sigma_{CO}$ with the new estimate of $\alpha_{CO}$ until we get a value of $\alpha_{CO}$ that does not vary by more than 0.1.
We obtain a mean Toomre Q of $0.94\pm0.48$, $0.38\pm0.21$, and $0.48\pm0.24$ for methods (1), (2) and (3), respectively. The error is the standard deviation.

Typical values for the local Universe are higher than one and suggest that galaxies have a stable disk dominated by pressure and rotation. A value of $Q\sim1$, or $Q\sim0.7$ for thick disks \citep{Kim2007, Dekel2009, Romeo2011} expected in turbulent galaxies at high redshift  \citep{Genzel2011, Elmegreen2017, Girard2019}, indicates a quasi-stable disk, where the gravity starts dominating and could cause the fragmentation of the disk.  The values derived from the ionized gas in this work agree with galaxies with a quasi-stable disk ($Q_{gas}\sim0.7-1$), similarly to the recent work from \citet{White2017}. Simulations agree with this picture by predicting a value of $Q\sim1$ at high-redshift and $Q\sim2-3$ in the local Universe \citep{Kim2007, Hopkins2011, Ceverino2010}. \citet{Meng2019} found slightly lower values of $Q\sim0.5-1$. Smaller values than unity ($Q_{gas}\sim0.6-0.8$) have been found in clumpy galaxies using the ionized gas as a kinematic tracer and have been interpreted as clumps being the results of an unstable disk \citep{Fisher2014}.
Values as low as 0.2 have been observed in star-forming clump regions \citep{Genzel2011, Genzel2014, Fisher2017a}.

Our values derived from the molecular gas dispersion are lower than the theoretical predictions, which suggest that galaxies self-regulate with $Q\sim0.7-1$  \citep[][]{Dekel2009, Krumholz2018}. A Toomre $Q$ parameter lower than one (or $\sim0.7$ for a thick disk) indicates unstable disks and that galaxies should collapse due to gravitation or that gravitational instabilities are not the main effect regulating galaxies. When using a variable $\alpha_{CO}$ with a $\sigma_{0,mol}$ the derived values of $Q$ are still low, but the scatter is within the range of the critical value of 0.7 for thick disks.

\section{Effect of using molecular gas dispersion on comparisons to star formation theories}
\label{sec:4}

In this section, we explore the relation between the velocity dispersion and star formation rate in addition to the relation between the star formation rate surface density and pressure. We also compare our observations to simulations and galaxy evolution and stellar feedback models. 

\subsection{Correlation of  $\sigma_0$ with star formation rate}
\label{sec:sfr}

We explore the relationship between the velocity dispersion and the star formation rate in Fig. \ref{fig:disp_sfr}. We find a strong correlation with a Spearman coefficient of 0.71 for the molecular gas. This means that the most gas-rich, turbulent disk galaxies of our sample are also more star-forming. A relation between $\sigma_0$ and SFR could be seen if our sample selection is biased to the brightest objects at high redshift  and to less star-forming galaxies in the local Universe since highly star-forming local galaxies are rare. However, this is unlikely since our sample includes lensed main-sequence galaxies at $z\sim1$ with lower SFR \citep{Girard2019} and nearby galaxies from DYNAMO with a higher SFR \citep{Fisher2017a}. These galaxies fall very well on the relation and show similar dispersion as other galaxies with the same SFR. 
This also suggests that the processes that convert the gas into stars are independent of redshift.

One scenario that has been suggested to explain the turbulence in the disk is the star formation feedback, which includes all the energy that is released in the interstellar medium from the stellar formation (stellar winds, supernovae, etc.). It is now well established that feedback only models, with traditional assumptions about feedback efficiency, do not produce a strong correlation between the star formation rate and the dispersion. \citet{Ostriker2011} and \citet{Shetty2012} predict values between $\sigma_0 \sim3-12$ \kms \, that barely vary with SFR. \citet{Krumholz2018} found similar results for a model with fixed star formation efficiency per free fall time, $\epsilon_{ff}$ (see Fig. \ref{fig:disp_sfr}, left panel). Some models also predicted higher velocity dispersion of $\sigma_0>50$ \kms, when including non-standard assumptions, such as boosting of feedback from clustering of supernova \citep{Gatto2015, Hopkins2011, Gentry2017, Fielding2018}. Overall, with fiducial assumptions, the models predict that the maximum velocity dispersion that the feedback can sustain is about 15 \kms. Even though the molecular gas dispersion is lower, it is still true that fixed-efficiency feedback-only models do no produce high enough dispersions. 

\citet{Krumholz2018} suggested that the discrepancy at high star formation rate could be alleviated by gravitational instabilities (or transport). In this model, the turbulence in the rotating galaxies depends on two main drivers, the feedback and the energy loss from gravitationally driven gas flows. They found that this model predicts values consistent with the ionized gas velocity dispersion obtained in several surveys. \citet{Ubler2019} also found that their ionized gas observations from KMOS$\rm^{3D}$ are in good agreement with this model. In this work, we find that taking the recommended prescriptions for those models over predicts the molecular gas velocity dispersion, as seen in Fig. \ref{fig:disp_sfr} (right panel). The SFR of high-redshift galaxies is a factor $\sim10$ higher than the predictions from the model at a fixed dispersion. We note here that this result is obtained from the SFR and velocity dispersion, two quantities that do not depend on the CO-to-H$_2$ conversion factor, $\alpha_{CO}$.

\citet{FaucherGiguere2013} and \citet{Krumholz2018} also presented a stellar feedback model with $\epsilon_{ff}$ free to vary (see Fig. \ref{fig:disp_sfr}, left panel). This model predicts an increase of the velocity dispersion with SFR with a dependence in SFR$\propto \sigma^2$. This model poorly reproduces values from the ionized gas, though it does seem to model the molecular gas relation quite well. Overall, the predicted velocity dispersion from the model are slightly higher than the molecular gas velocity dispersion of the most star-forming galaxies of our sample, but this model better describes our observations from the molecular gas than models assuming fixed $\epsilon_{ff}$ or including gravitational instabilites.

\begin{figure}[t]
\centering
\includegraphics[width=\columnwidth]{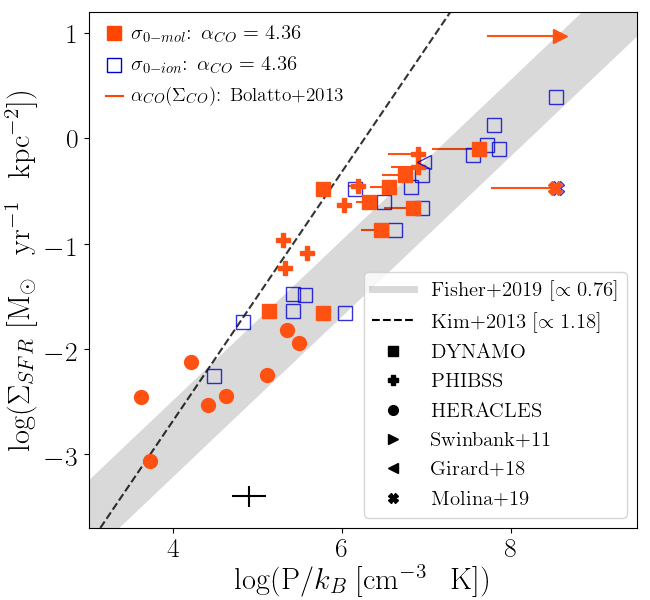}
\caption{SFR surface density as a function of the midplane hydrostatic pressure, $P_H$. 
We include the measurements from the same sample as presented in previous figures when the different quantities were available. The measurements obtained using the molecular and ionized gas dispersion are indicated with dark orange and blue empty symbols, respectively. The dashed lines indicates the relationship obtained from \citet{Kim2013} using simulations, and the shaded area indicates the fits obtained by \citet{Fisher2019} using ionized gas dispersion. The typical error on the values is indicated at the bottom. The symbols and orange lines indicate the values of $P_{H}$ obtained assuming $\alpha_{CO}=4.36$ and $\alpha_{CO}(\Sigma_{CO})$ using eq. 31 from \citet{Bolatto2013}, respectively.}

\label{fig:pressure}
\end{figure}

\subsection{Correlation of $\Sigma_{SFR}$ with pressure}

\label{sec:pressure}

In this section, we investigate the relationship between the SFR surface density, $\Sigma_{SFR}$, and pressure in the galaxies. We report in this work new pressure measurements derived using the molecular gas velocity dispersion for high-pressure systems. A correlation between these two quantities have been predicted by theoretical models and simulations \citep{Ostriker2011,Kim2013}. Observations have also shown that there is a clear correlation \citep{Leroy2008, Herrera-Camus2017, Sun2020}, but the observed relation does not always agree with predictions from the models, especially for the most star-forming galaxies in which the pressure is high \citep{Fisher2019}. A correlation between these two properties suggests that the pressure in galaxies is regulated by stellar feedback processes.

First we estimate the total midplane hydrostatic pressure using the equation from \citet{Elmegreen1989}:

\begin{equation}
\label{eq:PH}
P_H = \frac{\pi G}{2} \Sigma_g \left( \Sigma_g + \frac{\sigma_g \Sigma_\star}{\sigma_\star} \right)
\end{equation}
where $\Sigma_g$ and $\Sigma_\star$ are the gas (molecular and neutral) and stellar mass surface densities and $\sigma_g$ and $\sigma_\star$ are the gas and stellar velocity dispersion.

We also estimate the dynamical equilibrium pressure using the following equation \citep{Kim2011}:

\begin{equation}
\label{eq:PDE}
P_{DE} \approx \frac{\pi G}{2} \Sigma_g^2  + \Sigma_g ( 2 G \rho_{SD} )^{1/2} \sigma_g
\end{equation}
where the total midplane density $\rho_{SD}= \Sigma_\star/(4 h_z) + (\upsilon_{rot} / R_{disk})^2 / (4 \pi G)$  \citep{Leroy2008, Ostriker2011}. The first term represents the stellar component, where $h_z$ is the stellar scale height, and the second term represents the contribution from dark matter. 

To obtain $\Sigma_{SFR}$, $\Sigma_{mol}$ and $\Sigma_\star$ for the DYNAMO galaxies, we derive the half-light radius, $R_{1/2}$, by fitting an exponential profile to the H$\alpha$, CO(3-2) or CO(4-3), and F125W/HST maps when available. We obtain that $R_{1/2, H_\alpha}$ = ($1.1\pm0.2$) $R_{1/2, CO}$ and $R_{1/2, H_\alpha}$ = ($0.8\pm0.2$) $R_{1/2, \star}$ in the DYNAMO sample. To estimate the total gas mass density, $\Sigma_g$, we assume a HI gas surface density of 5 \msun $\rm pc^{-2}$ \citep{Bigiel2008}. We find that assuming a larger value of 15 \msun $\rm pc^{-2}$ could lead to a difference of 0.1 dex for the pressure in the DYNAMO galaxy with the lower pressure. The difference is typically 0.02-0.03 dex for the other DYNAMO galaxies. This is taken into account in the error.

In Fig. \ref{fig:pressure}, we also present measurements for other galaxies from the same sample shown in Figs. \ref{fig:disp_fgas} and \ref{fig:disp_sfr}. For the HERACLES (or THINGS) galaxies, we derive the HI gas mass density from the HI masses available in \citet{Leroy2008}. We also use $R_{1/2,SFR}$, $R_{1/2, CO}$ and $R_{1/2, \star}$ for each corresponding surface density and the molecular velocity dispersion derived in \citet{Mogotsi2016}. For the high-redshift galaxies presented in this figure, we assume the same HI gas surface density of 5 \msun $\rm pc^{-2}$ used for the DYNAMO galaxies and that $R_{1/2,SFR} \approx R_{1/2, CO} \approx R_{1/2, \star}$ as found in several studies \citep[e.g.][]{Tacconi2013, Bolatto2015}. We note that assuming $R_{1/2, H_\alpha} = 1.1 \, R_{1/2, CO}$ and $R_{1/2, CO} = 0.8 \, R_{1/2, \star}$, as found in this work for DYNAMO galaxies, could increase the pressure by a factor 1.5-2.0. (0.2-0.35 dex) for the PHIBSS galaxies.

To determine the stellar dispersion, we use the approximation $\sigma_\star \approx 0.5 \sqrt{\pi G l_\star \Sigma_\star}$, where $l_\star$ is the disk scale length. The disk scale length can be determined from $l\star=R_{1/2, \star}/1.76$. We find a good agreement between the values obtained from this estimation with the values derived for two DYNAMO galaxies by \citet{Bassett2014}. We also use the equation $\sigma_{\star} = 2.45 \times \sigma_{0,mol} + 15$ to estimate the stellar velocity dispersion \citep{Fisher2019}. We find a maximum difference of $\sim0.05$ dex for the pressure using this method.

To determine the dynamical equilibrium pressure, $P_{DE}$, one needs to assume the stellar scale height, $h_z$, of the galaxies. Local galaxies typically have a scale height of $\sim100$ pc while high-redshift galaxies have a thicker disk of $500-1000$ pc \citep{Elmegreen2005, Bassett2014, Elmegreen2017}. \citet{Elmegreen2017} found on average a scale height of $630\pm240$ pc for galaxies at $z\sim2$. In this work, we therefore assume a value of $100$ pc and $630$ pc for local and high-redshift or DYNAMO galaxies, respectively. We find that assuming 500 pc and 1000 pc instead of 630 pc for the high-redshift and DYNAMO galaxies could lead to differences of 0.02 dex. We note that our results suggest that the molecular gas disk thickness is much lower than the ionized and stellar disk thickness.

Fig. \ref{fig:pressure} presents the midplane hydrostatic pressure measurements from the ionized gas (blue empty symbols) and molecular gas (dark orange symbols). We find that at high $\Sigma_{SFR}$, pressure derived using molecular gas velocity dispersion is still on average an order of magnitude larger than expectations from theory and simulation. We find median and mean differences between the values measured from the ionized and molecular gas of the DYNAMO galaxies of 0.24 dex and 0.22 dex, respectively. This means that pressure measurements derived using the ionized gas as a tracer can be over-estimated by about $\sim0.22$ dex. We only show $P_H$ in Fig. \ref{fig:pressure} since $P_H$ and $P_{DE}$ are very similar (difference of $0.03$ dex and $0.06$ dex for DYNAMO and the whole sample, respectively). The values are presented in Table \ref{tab:appendix}.

We find a strong correlation with a Spearman coefficient of 0.92 for the molecular gas. This suggests that the pressure in both low and high SFR density environments is regulated by stellar feedback processes. We fit a linear relation using the molecular dispersion measurements and obtain a slope of $0.70\pm0.07$. We also show values of $P_{H}$ assuming that $\alpha_{CO}$ varies with $\Sigma_{CO}$ in Fig. \ref{fig:pressure} (orange line). As described previously in Sect. \ref{sect:toomreQ}, we derive $\alpha_{CO}(\Sigma_{CO})$ following eq. 31 from \citet{Bolatto2013}. We find that the pressures are lower by 0.21 dex on average for the DYNAMO galaxies. We fit again a linear relation using the whole sample and obtain a slope of $0.83\pm0.08$, consistent with the results previously obtained by \citet{Fisher2019} of $0.76\pm0.06$ with a similar sample. \citet{Sun2020} also found a sublinear slope of $0.84\pm0.01$ from their observations of local galaxies at kpc-scales.
This means that observations are in agreement with sublinear slopes and are lower than predictions from simulations and theoretical derivations, which estimate a slope of $\sim1-1.2$ \citep{Ostriker2010, Kim2011, Kim2013, Muratov2015}. This indicates that the pressure in high SFR density galaxies is higher than predicted by feedback models.

\citet{Fisher2019} suggested that this could be due to the variation of the feedback momentum injected into the interstellar medium, $p_*/m_*$, with the local environments. One possible candidate is that supernova clustering leads to more efficient feedback \citep{Gentry2020}. Alternatively, mechanisms other than feedback from stars could also drive turbulence in the most star-forming galaxies. \citet{Krumholz2018} suggested that gravitational instabilities (or transport), in addition to feedback, could drive more turbulence and lead to higher pressure. However, from Fig. \ref{fig:disp_sfr} (right panel) we find that this model overestimates the molecular gas velocity dispersion observed in the most star-forming galaxies.

\section{Summary}
\label{Sect:conclusion}

We present resolved ALMA and GMOS/Gemini observations of 9 galaxies from DYNAMO, a sample of rare galaxies at $0.075<z<0.2$ showing similar physical and kinematic properties to $z\sim1-2$ main-sequence galaxies. We combine our sample with 12 galaxies at $z\sim0.5-2.5$ and find that the molecular gas velocity dispersion is systematically lower by a factor of $2.45\pm0.38$ compared to the ionized gas velocity dispersion, after correcting for thermal broadening and HII region expansion. This difference does not depend on gas fraction and indicates the co-existence of a thin molecular gas disk and thick ionized gas disk in these galaxies. The nature of this difference remains unclear. One possible explanation could be the presence of fast shocks that dissociate the molecular gas.  

We find a strong correlation between $\sigma_{0}$ and $f_{gas}$ and obtain a fit of log($\sigma_{0,mol}/$\kms) = $0.87 \times f_{gas} + (0.90\pm0.03)$ and log($\sigma_{0,ion\text{-}th}/$\kms) = $0.87 \times f_{gas} + (1.29\pm0.03)$, with a zero-offset of $0.39\pm0.06$ dex between the ionized and molecular gas. We also find that $\sigma_{0,mol}$ is higher in high-redshift galaxies and show that the EAGLE and FIRE simulations overall predict higher values. TNG50 better predicts the observations. 
We obtain Toomre Q values of $0.94\pm0.48$, $0.38\pm0.21$, and $0.48\pm0.24$ on average for the DYNAMO sample when assuming $\alpha_{CO}=4.36$ and $\sigma_0=\sigma_{0,ion}$, $\alpha_{CO}=4.36$ and $\sigma_0=\sigma_{0,mol}$, and $\alpha_{CO}(\Sigma_{CO})$ and $\sigma_0=\sigma_{0,mol}$, respectively. Our estimations using the molecular gas velocity dispersion are typically lower than predictions from gravitational instability theory of $Q\sim0.7-1.0$.

We obtain a strong correlation between the molecular velocity dispersion and the star formation rate. When comparing our results to predictions from analytical star formation models, we find that stellar feedback models assuming a constant $\epsilon_{ff}$ are not able to reproduce the higher molecular gas velocity dispersion of the more star-forming galaxies. We also find that models that include both stellar feedback and gravitational instabilities (or transport), and that were previously found to predict well the ionized gas velocity dispersion, overestimate the molecular velocity dispersion. However, stellar feedback models with $\epsilon_{ff}$ free to vary predict values similar to our observations.

We explore the correlation between SFR surface density, $\Sigma_{SFR}$, and pressure and find that both the total midplane hydrostatic pressure, $P_H$, and dynamical equilibrium pressure, $P_{DE}$, derived using the molecular gas velocity dispersion are lower by  $\sim0.22$ dex compared to the values derived using the ionized gas velocity dispersion. When assuming a variable $\alpha_{CO}$, we find that the pressure is lower by $\sim0.21$ dex in the DYNAMO galaxies.
We find a sublinear relation of $0.70\pm0.07$ and $0.83\pm0.08$ for $\alpha_{CO}=4.36$ and a variable $\alpha_{CO}(\Sigma_{CO})$, in agreement with previous results from \citet{Fisher2019}, but lower than predictions from stellar feedback models. 

The disagreement between predictions from feedback models and the pressure and turbulence observed in the DYNAMO galaxies and other high-redshift galaxies could indicate that another mechanism drives these quantities. Another possibility is that the star formation efficiency, $\epsilon_{ff}$, and/or the feedback efficiency, $p_*/m_*$, vary in galaxies, and are higher in the densest environments. Clearly, more studies on the molecular gas kinematics in high-redshift galaxies are needed to better understand the nature of the turbulence and high pressure in these galaxies.

\acknowledgments
MG and DBF acknowledge support from Australian Research Council (ARC) DP grant DP160102235 and Future Fellowship FT170100376. 
RB acknowledges support from the Australian Research Council Centre of Excellence for All Sky Astrophysics in 3 Dimensions (ASTRO 3D) through project number CE170100013.
EJ acknowledges the support of the University of Western Australia through a Scholarship for International Research Fees and a Co-Funded Postgraduate Award.
DO is a recipient of an Australian Research Council Future Fellowship (FT190100083) funded by the Australian Government.
We are grateful to Hannah \"Ubler and the PHIBSS team for sharing information. 
This paper makes use of the following ALMA data: ADS/JAO.ALMA$\#$2017.1.00239.S and ADS/JAO.ALMA$\#$ 2019.1.00447.S. ALMA is a partnership of ESO (representing its member states), NSF (USA) and NINS (Japan), together with NRC (Canada), MOST and ASIAA (Taiwan), and KASI (Republic of Korea), in cooperation with the Republic of Chile. The Joint ALMA Observatory is operated by ESO, AUI/NRAO and NAOJ.

\facilities{ALMA, Gemini} 

%






\newpage

\appendix
\section{Kinematic maps}

Figure \ref{map_galaxies} shows the HST F125W images, the flux maps, the observed velocity maps and the observed velocity dispersion maps of ALMA and GMOS/Gemini data of each galaxy from the DYNAMO sample. We also present the observed rotation curves, the modeled rotation curves and the dispersion profiles of each galaxy.

\begin{figure*}
\includegraphics[scale=1.]{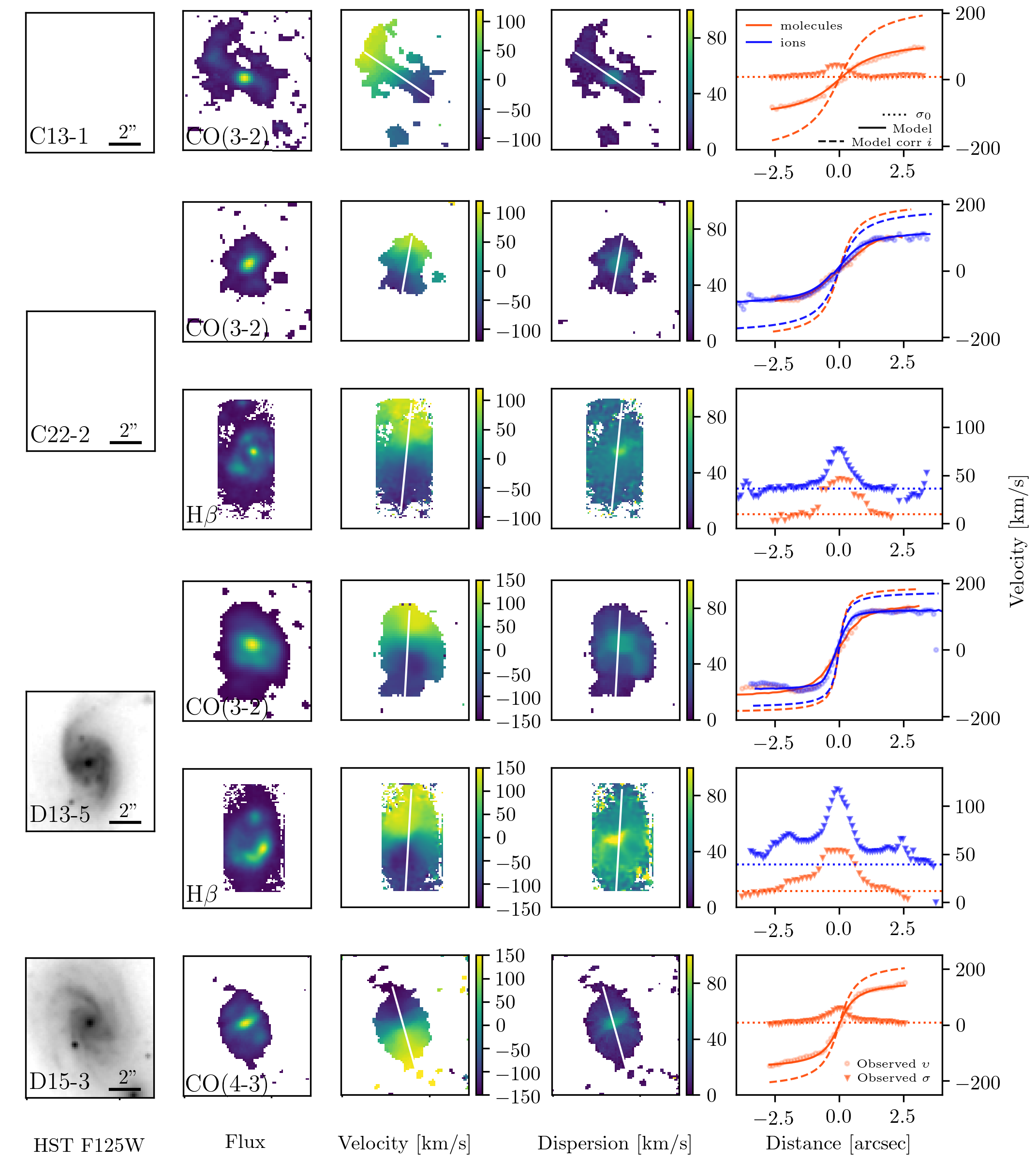}
%
\caption{HST F125W images, the flux maps, the observed velocity maps and the velocity dispersion maps of each galaxy from the DYNAMO sample. The white lines in the velocity and dispersion maps indicate the kinematic major axis. In the right panels, we present the rotation curves (circle) and dispersion profiles (triangle) extracted on the major axis of the observed velocity and dispersion maps and the rotation curves extracted on the major axis of the velocity maps from the kinematic models (solid line). The intrinsic rotation curves corrected for inclination and intrinsic velocity dispersion from the models are indicated with a dashed line and dotted line, respectively. The molecular and ionised gas are in orange and blue, respectively.}
\label{map_galaxies}
\end{figure*}

\begin{figure*} 
\centering
\setcounter{figure}{5}
\includegraphics[scale=1]{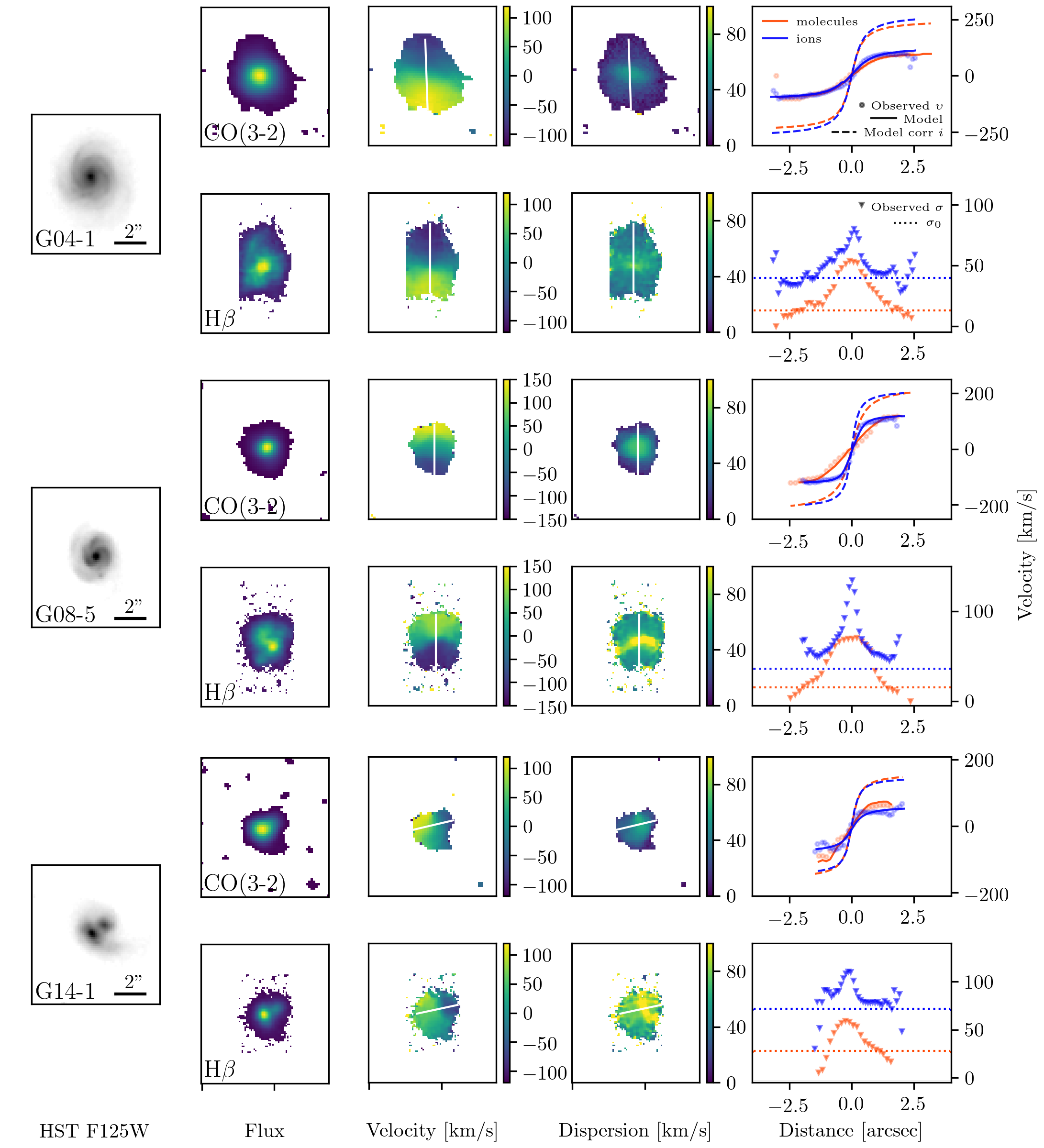}
\caption{Continued.}
\end{figure*}

\begin{figure*} 
\centering
\setcounter{figure}{5}
\includegraphics[scale=1]{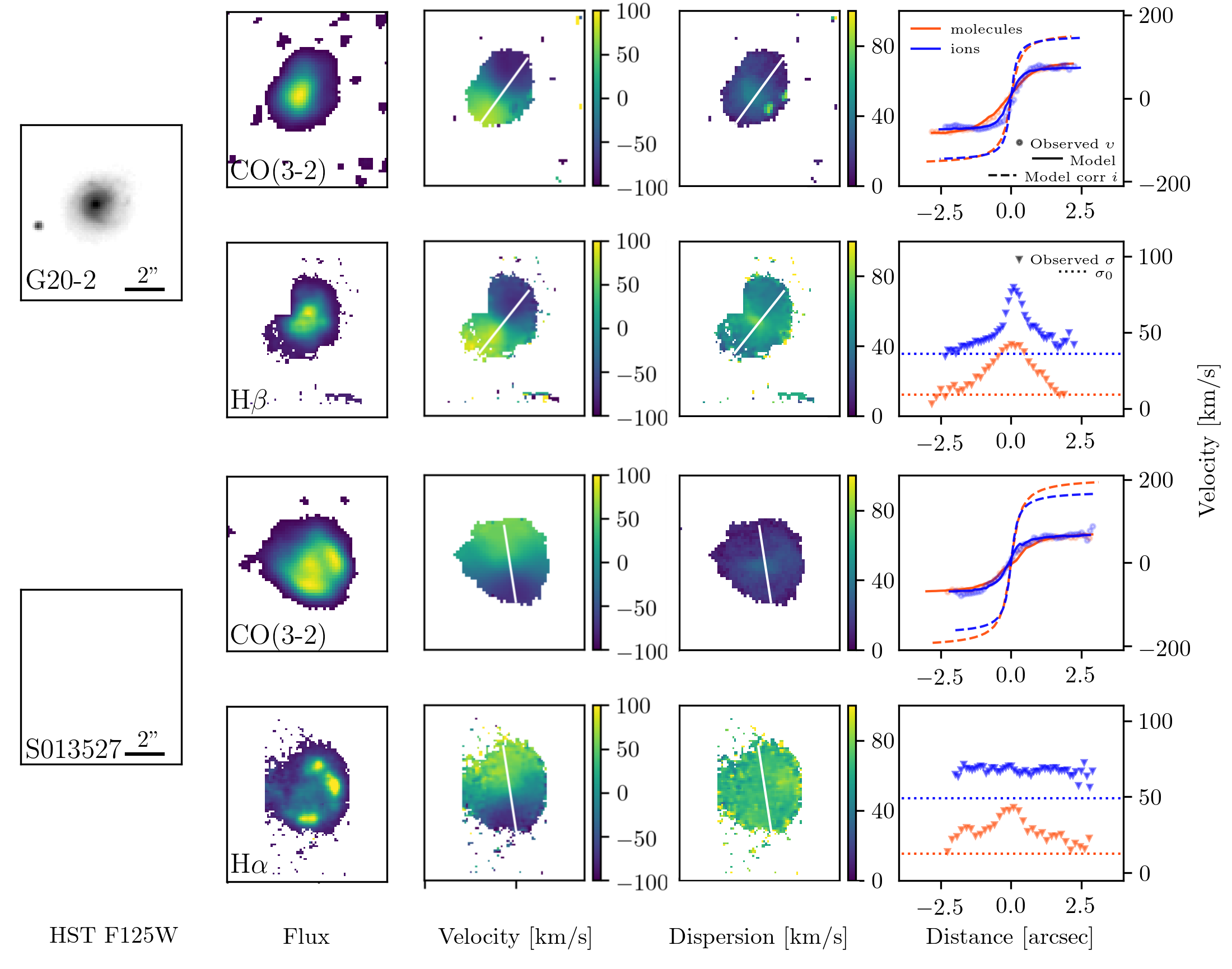}
\caption{Continued.}
\end{figure*}

\section{Galaxy properties}

Table \ref{tab:appendix} presents the  H$\alpha$, CO and stellar half-light radii derived for the DYNAMO sample. We also show the values from the literature for the high-redshift and local galaxy samples. We show the derived values of $P_H$ using the ionised and molecular gas velocity dispersion assuming $\alpha_{CO}=4.36$, $P_H(\alpha_{CO})$ using the molecular gas velocity dispersion and assuming a variable $\alpha_{CO}(\Sigma_{CO})$ following the prescription of \citet{Bolatto2013}, and $P_{DE}$ using the molecular gas velocity dispersion and $\alpha_{CO}=4.36$.
\newpage

\begin{deluxetable*}{lccccccccc}
\caption{Galaxy properties\label{tab:appendix}}
\tablewidth{0pt}
\tablehead{
Galaxy & \colhead{R$_{1/2,SFR}\, ^{a}$ } & \colhead{R$_{1/2,CO}\, ^{a}$} & \colhead{R$_{1/2,\star}\, ^{a}$}  & \colhead{P$_{H,ion}/k_B$} & \colhead{P$_{H,mol}/k_B$} & \colhead{P$_{H,mol}(\alpha_{CO})/k_B$} & \colhead{P$_{DE,mol}/k_B$} & \colhead{Reference for the radius }
\\
\colhead{} & \colhead{[kpc]} & \colhead{[kpc]} & \colhead{[kpc]} & \colhead{[log(cm$^{-3} \, K$)]} & \colhead{[log(cm$^{-3} \, K$)]} & \colhead{[log(cm$^{-3} \, K$)]} & \colhead{[log(cm$^{-3} \, K$)]} & \colhead{}
}
\startdata
C13-1  & 4.2 & 3.3 & -  &   5.42 $\pm 0.22$    &  5.13$\pm 0.24$ &   5.12$\pm 0.24$   &  5.10$\pm 0.25$ & This work \\
C22-2  & 3.4 & 1.9 & - &    6.03 $\pm 0.14$   &  5.78$\pm 0.15$ &     5.76$\pm 0.15$  &   5.75$\pm 0.17$ & This work\\
D13-5  & 2.0 & 1.9 & 2.5 &   6.82  $\pm 0.13$   &  6.56$\pm 0.12$  &   6.34$\pm 0.13$   &  6.50$\pm 0.16$  & This work   \\
D15-3  & 2.2 & 1.7 & 3.0 &    6.63 $\pm 0.12$    & 6.47$\pm 0.11$ &     6.23$\pm 0.11$  &  6.44$\pm 0.13$ & This work\\
G04-1  & 2.8 & 2.4 & 3.3 &    6.95 $\pm 0.13$     & 6.84$\pm 0.12$ &    6.49$\pm 0.12$  &   6.82$\pm 0.14$  & This work \\
G08-5  & 1.8 & 1.7 & 2.3 &   6.50$\pm 0.15$     & 6.32$\pm 0.14$ &       6.17$\pm 0.14$ &    6.28$\pm 0.15$ & This work  \\
G14-1  & 1.1 & 1.1 & 2.7 &    6.95$\pm 0.15$     & 6.75$\pm 0.13$ &     6.48$\pm 0.13$   &   6.73$\pm 0.15$  & This work     \\
G20-2  & 2.1 & 2.1 & 2.2  &    6.15 $\pm 0.16$     & 5.77$\pm 0.18$ &     5.77$\pm 0.18$   &   5.69$\pm 0.19$  & This work   \\
SDSS 013527-1039 & 1.6 & 1.7 & - & 7.87$\pm 0.14$  & 7.62$\pm 0.12$ &     7.07$\pm 0.12$   &    7.56$\pm 0.13$  & This work  \\
\hline
SMM J2315-0102  & - & $1.5$ & - & - & 8.58$\pm 0.19$ &  7.72$\pm 0.17$ & 8.55$\pm 0.20$ & \citet{Swinbank2011} \\
Cosmic Snake  & 6.0$\pm1.2$ & 0.75$\pm0.03$ & - & 7.84$\pm 0.16$  & 7.84$\pm 0.18$ & 7.09$\pm 0.17$ & 7.85$\pm 0.19$ & \citet{Girard2019} \\
SHiZELS-19 & 1.80$\pm0.16$ &  1.68$\pm0.03$ & - & 8.54$\pm 0.14$  & 8.52$\pm 0.16$ & 7.77$\pm 0.14$  & 8.46$\pm 0.17$ & \citet{Molina2019} \\
G3\_10098     & - & - & 8.5 & - & 5.33$\pm 0.28$  & 5.32$\pm 0.28$ & - & PHIBSS$\, ^{b}$ \\
G4\_21351      & - & - & 7.9 & - & 5.30$\pm 0.27$  & 5.30$\pm 0.27$ & - & PHIBSS$\, ^{b}$ \\
EGS\_13035123 & - & - &  9.1 & - & 5.59$\pm 0.27$  & 5.58$\pm 0.27$ & 5.62$\pm 0.28$ & PHIBSS$\, ^{b}$ \\
EGS\_12007881 & - & - & 6.0 & - & 6.02$\pm 0.20$  & 5.94$\pm 0.20$ & 6.02$\pm 0.20$ & PHIBSS$\, ^{b}$ \\
EGS\_13003805 & - & - & 5.6  & - & 6.91$\pm 0.20$  & 6.58$\pm 0.19$  & 6.90$\pm 0.20$  & PHIBSS$\, ^{b}$  \\
EGS4\_24985   & - & - &  4.9 &- & 6.19$\pm 0.19$  & 6.08$\pm 0.19$  & - & PHIBSS$\, ^{b}$  \\
EGS\_13011166 & - & - & 6.9  & - & 6.90$\pm 0.20$  & 6.54$\pm 0.19$  & 6.90$\pm 0.21$ & PHIBSS$\, ^{b}$  \\
\hline
NGC628 & 4.2 & 4.2 & 4.0 & - & 4.62$\pm 0.15$  & 4.62$\pm 0.15$ & 4.76$\pm 0.16$ & \citet{Leroy2008}\\ 
NGC925 & 7.2 & -  & 7.2  & - & 3.72$\pm 0.15$  & 3.72$\pm 0.15$ & 3.92$\pm 0.16$ & \citet{Leroy2008}\\ 
NGC2976 & 1.4 & 2.1 & 1.6 & - & 3.62$\pm 0.19$  & 3.62$\pm 0.19$ & 3.71$\pm 0.20$ & \citet{Leroy2008}\\ 
NGC3184 & 4.9 & 5.1 & 4.2 & - & 4.41$\pm 0.16$  & 4.41$\pm 0.16$ & 4.61$\pm 0.16$ & \citet{Leroy2008}\\ 
NGC3351 & 3.2 &  4.4 & 3.9 & - & 4.21$\pm 0.18$  & 4.21$\pm 0.18$ & 4.42$\pm 0.19$ & \citet{Leroy2008}\\ 
NGC4736 & 1.6 & 1.4  & 1.9 & - & 5.35$\pm 0.18$  & 5.35$\pm 0.18$ & 5.44$\pm 0.18$ & \citet{Leroy2008}\\ 
NGC5055 & 5.5 & 5.5 & 5.6 & - & 5.11$\pm 0.16$  & 5.11$\pm 0.16$ & 5.29$\pm 0.17$ & \citet{Leroy2008}\\ 
NGC6946 & 4.8 & 3.4 & 4.4 & - & 5.50 $\pm 0.15$ & 5.49$\pm 0.15$ & 5.61$\pm 0.15$ & \citet{Leroy2008}\\ 
\enddata
\tablecomments{\\
$^{a}$ Typical uncertainty on the radius of the DYNAMO galaxies is 15\%. We assume a similar error on the radius of other galaxies when the uncertainty was unavailable. \\
$^{b}$ PHIBSS refers to \citet{Tacconi2013}, \citet{Tacconi2018}, \citet{Genzel2015}, and \citet{Freundlich2019}. The radius was measured in the rest-frame optical. 
}
\end{deluxetable*}  

\bibliography{reference}{}
\bibliographystyle{aasjournal}



\end{document}